\documentclass[showpacs,preprintnumbers,amsmath,amssymb,twocolumn]{revtex4}
\usepackage{graphicx}   
\usepackage{dcolumn}    
\usepackage{bm}         

\begin{document}

\title{Quantum information processing using strongly-dipolar coupled nuclear spins}

\author{T. S. Mahesh and Dieter Suter}
\address{Department of Physics, University of Dortmund, 44221 Dortmund, Germany}

\date{\today}

\begin{abstract}
{Dipolar coupled homonuclear spins present challenging, yet useful
systems for quantum information processing. 
In such systems, eigenbasis of the system Hamiltonian
is the appropriate computational basis and coherent control can
be achieved by specially designed strongly modulating pulses.
In this letter we
describe the first experimental implementation of the quantum 
algorithm for numerical gradient estimation on the eigenbasis of
a four spin system.}
\end{abstract}

\maketitle

An important issue in experimental quantum information processing (QIP) is 
achieving coherent control while increasing the number of qubits
\cite{divincenzo,corychuangrev}.  
In all existing implementations of quantum computation, the execution time
is limited by durations of nonlocal gates. 
In the case of nuclear magnetic resonance (NMR) implementations of QIP,
where qubits are formed by mutually coupled spin 1/2 nuclei,
most of the existing implementations used isotropic liquids.
In these systems, the durations of nonlocal gates are limited by the strength
of (indirect) scalar couplings, which typically are of the order of 10-100 Hz. 

Apart from the scalar couplings, nuclear spins also interact through magnetic
dipole-dipole couplings, which are 2-3 orders of magnitude stronger than the
scalar couplings, and would therefore allow significantly faster gate operations.
In the case of isotropic liquids, the rapid molecular reorientation averages 
the dipolar interactions to zero. 
On the other hand, in oriented systems, the dipolar
interactions survive and are therefore potentially better candidates for NMR-QIP
\cite{cory3qss}.  

The Hamiltonian for a dipolar coupled $n$-spin system is,
\begin{eqnarray}
{\cal H} = \sum_{j=1}^n \hbar \omega_j I^j_z + 
\sum_{j=1,k<j}^{n} 2\pi\hbar {\mathrm D}_{jk}
(3{\mathrm{I}}_z^j \cdot {\mathrm{I}}_z^k - 
{\mathrm{\bf I}}^j \cdot {\mathrm{\bf I}}^k) 
\end{eqnarray}
where $\omega_j$ are the chemical shifts, ${\mathrm D}_{jk}$ 
are the dipolar coupling constants and 
$I_z^j$ are z-components of the spin angular 
momentum operators {\bf I}$^j$ \cite{ernstbook}.
The first term corresponds to the Zeeman interaction and the 
second term describes the dipolar interaction.
When $\vert \mathrm{D}_{jk} \vert \ll \vert \omega_j - \omega_k \vert$,
one usually invokes the `weak-coupling' approximation \cite{ernstbook,levittbook},
where one ignores the off-diagonal parts of the Hamiltonian.  
Then the Zeeman product states are
the eigenstates of the full Hamiltonian and 
individual spins can be conveniently treated as qubits.  
Most of the
experiments in NMR-QIP have so far been carried out on such systems
\cite{corychuangrev}.  

However to maximize the
execution speed of our quantum processor it is desirable to
use stronger couplings, including
$\vert \mathrm{D}_{jk} \vert \ge \vert \omega_j - \omega_k \vert$.
In this case, we have to use the complete Hamiltonian (1), where
the Zeeman and the coupling parts do not commute and not all eigenstates
are Zeeman product states but linear combinations of them.  
Unlike in the case of a weakly coupled spin system,
individual spins of a strongly coupled system loose addressability.
In such cases the eigenstates of the full Hamiltonian can
form individually controllable subsystems and then the eigenbasis
becomes the natural and accurate choice for the computational basis.

Though strongly dipolar coupled spin-systems have been suggested for QIP earlier
\cite{maheshCS}, so far it had not been possible to implement general unitary
gate operations that can be applied to arbitrary initial conditions.
Here we show how quantum computation can be realized on the eigenbasis
by efficient coherent control techniques. 
As a specific system, we use a strongly dipolar coupled four-spin 
system partially oriented in a liquid crystal.  Such systems have certain key merits.
Unlike in the liquid state systems, the intramolecular dipolar couplings are not
averaged out completely, but are only scaled down by the order parameter of the
solute molecules oriented in the liquid crystal \cite{diehlvol6,emsleybook}. 
Since intermolecular interactions are averaged out (unlike the crystalline systems),
liquid crystalline systems provide well defined quantum registers
with low decoherence rates.
We achieve high-fidelity coherent control with the help of specially designed strongly modulating pulses (SMPs).

To demonstrate these techniques, we use an interesting algorithm
suggested by S. P. Jordan \cite{spjordan}: 
the quantum algorithm for numerical gradient estimation (QNGE).
Before we discuss our implementation, we summarize
the theoretical description of QNGE \cite{spjordan}. 
The gradient of an one-dimensional real function $f$ over a small real range
$l$ is written as $\nabla f = \{f(l/2) - f(-l/2)\}/l$.
Thus classically two function evaluations are necessary to estimate the
gradient of a one-dimensional function and a minimum of $d+1$ function evaluations
are necessary for a $d$-dimensional function.  On the other hand, QNGE requires only one 
function evaluation independent of the dimension of the function.  
Here the 
function $f$ is encoded in an $n$-qubit input register.  
An ancilla
register is also required whose size ($n_0$ qubits) depends on
the maximum possible value of the gradient.

\begin{figure}
\includegraphics[bb=0 660 350 840,clip,width=7.5cm]{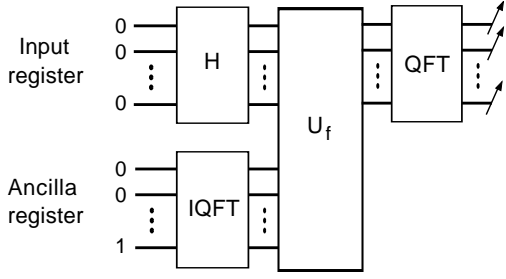}
\caption{\label{gecircuit}Circuit diagram for QNGE.}
\end{figure}

In QIP, numbers are represented in binary form.
To encode the real number $x$ in the input register, 
we have to convert it into a nonnegative integer $\delta \in \{0,1,\cdots, N-1\}$,
where $N = 2^n$.
The encoding is defined by
\begin{eqnarray}
x = \frac{l}{N-1}\left( \delta - \frac{N-1}{2} \right).
\label{eqnforx}
\end{eqnarray}
The circuit diagram for the quantum algorithm is shown in Figure 
\ref{gecircuit}.
Initially the ancilla register ($an$) is set to 1 and the input register 
($in$) is set to 0.  The inverse quantum Fourier transform (IQFT) 
prepares a plane wave state on the ancilla register and 
and the Hadamard transform (H) prepares a uniform superposition
on the input register \cite{chuangbook}:
\begin{eqnarray}
&&\underbrace{\vert 00 \cdots 01 \rangle}_{an}
\underbrace{\vert 00 \cdots 00 \rangle}_{in}
\stackrel{\mathrm{IQFT}^{an}}{\longrightarrow}
\stackrel{\mathrm{H}^{in}}{\longrightarrow} \nonumber \\
&&\hspace{2cm}\frac{1}{\sqrt{N_0}}
\sum_{k=0}^{N_0-1} e^{-i\frac{2 \pi k}{N_0}} \vert k \rangle
\frac{1}{\sqrt{N}}\sum_{\delta = 0}^{N-1}\vert \delta \rangle 
\label{afterqft} \nonumber
\end{eqnarray}
The ancilla register 
is now in an eigenstate of the addition modulo $N_0$.
On applying oracle 
$U_f: \vert an,in \rangle \rightarrow \vert sf(x) \oplus_{N_0} an,in \rangle$,
where $s$ is a scaling factor, the total state becomes
\begin{equation}
\frac{1}{\sqrt{N_0N}}
\sum_{k=0}^{N_0-1} e^{-i\frac{2 \pi k}{N_0}} \vert k \rangle
\sum_{\delta =0}^{N-1} e^{i \frac{2 \pi}{N_0} sf(x)} \vert \delta \rangle
\nonumber.
\label{afteraddition}
\end{equation}
For a small $x$, $f(x) \approx f(0)+x \nabla f$.  Substituting for $x$ from equation
\ref{eqnforx} and ignoring the global phase,
the input register reduces to,
\begin{eqnarray}
\frac{1}{\sqrt{N}}\sum_{\delta =0}^{N-1}
e^{i \frac{2 \pi}{N_0} \frac{ls \delta}{N-1} \nabla f}
\vert \delta \rangle \nonumber.
\end{eqnarray}
The scaling factor $s$ is set to maximize the precision i.e., 
\begin{eqnarray}
s = N_0 (N-1)/Nl.
\label{eqnfors}
\end{eqnarray}
Now a phase estimation is carried out with the quantum Fourier transform (QFT)
on the input register
\begin{eqnarray}
\frac{1}{\sqrt{N}}\sum_{\delta =0}^{N-1}
e^{i \frac{2 \pi \delta}{N} \nabla f}
\vert \delta \rangle
\stackrel{\mathrm{QFT}}{\longrightarrow}
\left\vert \nabla f \right\rangle \nonumber.
\label{phaseestim}
\end{eqnarray}
Measuring now the input registers in the computational basis gives an estimation
of $\nabla f$.  

In our 4-qubit case, we select two qubits as ancillas and the other two as input qubits i.e., $n_0 =n=2$,
$N_0=N = 4$, and $\delta \in 
\{\vert 00 \rangle, \vert 01 \rangle, \vert 10 \rangle, \vert 11 \rangle \}$.
From equation (\ref{eqnforx}), $x \in \{-1/2, -1/6, 1/6, 1/2\}$ for $l = 1$.  
Let us consider an example function $f(x) \in \{0, 2/3, 4/3, 2\} $,
which has a gradient $\nabla f = 2$.  Using equation \ref{eqnfors} we obtain 
the scaling factor $s=3$
so that $sf(x) \in \{0, 2, 4, 6\}$.  For $\delta \in 
\{\vert 00 \rangle,\vert 10 \rangle\}$, $sf(x)_{\oplus N_0}$ remains identity.
For $\delta \in \{\vert 01 \rangle,\vert 11 \rangle\}$, $sf(x)_{\oplus N_0}$
adds 2 to the ancilla, i.e., flips the first ancilla qubit.  Therefore,
for this example, the oracle is a CNOT($an_1,in_2$) gate i.e.,
a NOT operation on the first ancilla qubit 
controlled by the second input qubit.
The propagator for the entire algorithm is therefore,
\begin{eqnarray}  
U_{\mathrm{QNGE}} &=& U_{\mathrm{QFT}}^{in} \cdot U_{\mathrm{CNOT}(an_1,in_2)} 
\cdot U_{\mathrm{H}}^{in} 
\cdot U_{\mathrm{IQFT}}^{an} \nonumber.
\end{eqnarray}
We design a single SMP corresponding to this entire operation on
the input states.

\begin{figure}
\vspace{-0.6cm}
\includegraphics[width=7cm,angle=-90]{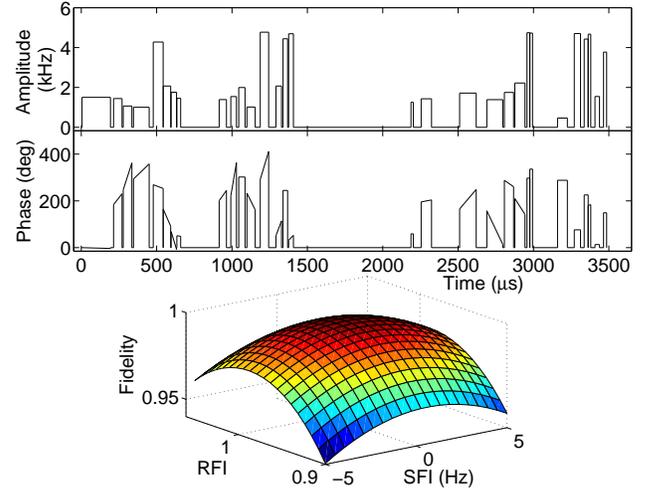}
\caption{\label{smprob} The SMP for QNGE:
(top) amplitude vs time, (middle) phase vs time, and (bottom)
fidelity vs RF inhomogeneity (RFI, 0.9 to 1.05 of the ideal field)  
and static field inhomogeneity (SFI, -5 Hz to 5 Hz).  The 30-segment SMP 
has an average fidelity $F'_{avg} = 0.981.$  
}
\end{figure}
An SMP is a cascade of radio-frequency (RF) pulses numerically calculated based on
the precise knowledge of the internal Hamiltonian of the qubit system and the target propagator
\cite{fortunato,navin}.
Given a target operator $U_{\mathrm T}$, 
Fortunato et al \cite{fortunato} have described numerically searching 
an SMP propagator $U_{\mathrm{SMP}}$ 
based on four RF parameters  for each segment $k$:
duration ($\tau^{(k)}$), amplitude ($\omega_A^{(k)}$), 
phase ($\phi^{(k)}$) and frequency ($\omega_F^{(k)}$).  
We found that it is useful to add one more degree of freedom:
after each pulse segment, we introduce a variable delay $\tau_d^{(k)}$.
The delays are computationally inexpensive to optimize, easy and accurate to implement
and make designing non-local gates easier.
All the parameters are determined so as to maximize the fidelity
\begin{eqnarray}
F = \mathrm{trace} \left[ U_{\mathrm T}^{-1} \cdot U_{\mathrm{SMP}} \right]/M \nonumber
\end{eqnarray} 
where $U_{\mathrm{SMP}}$ is the propagator for the SMP, 
$M$ is the dimension of the operators and $s$ is the number of segments.  
A MATLAB package has been developed
which uses the Nelder-Mead simplex algorithm as the maximization routine
\cite{maheshsmp}. 
The search constraints can be simplified if the input state
is definitely known.  The fidelity of an SMP specific to
a known initial state $\rho_{\mathrm{in}}$ 
can be written as,
\begin{eqnarray}
F' = \frac{\mathrm{trace} \left[ \rho_T \cdot \rho_{\mathrm{SMP}} \right]}
{\sqrt{\mathrm{trace} \left[ \rho_T^2 \right] \cdot \mathrm{trace} 
\left[ \rho_{\mathrm{SMP}}^2 \right]}},
\label{inspefid}
\end{eqnarray}
where $\rho_T = U_{\mathrm T} \cdot \rho_{\mathrm{in}} \cdot U_{\mathrm T}^{-1}$ and 
$\rho_{\mathrm{SMP}} = U_{\mathrm{SMP}} \cdot \rho_{\mathrm{in}} \cdot U_{\mathrm{SMP}}^{-1}$.
The SMPs are made robust against the spatial inhomogeneous
distributions of RF amplitudes and of static fields ($\Delta \omega_i$)
by maximizing an average fidelity, 
$F'_{avg} = \sum_i F'(\omega_{iA},\Delta\omega_i)$ \cite{nicrfi}.
Figure \ref{smprob} shows a single SMP performing the full QNGE
on the specific four-qubit system.
Though it is possible to decompose the target propagator 
into several SMPs each corresponding to one or two-qubit gates, 
it is more efficient, at least for small spin systems, to 
design and execute a single robust SMP implementing the 
entire algorithm.

\begin{figure}
\includegraphics[width=7cm]{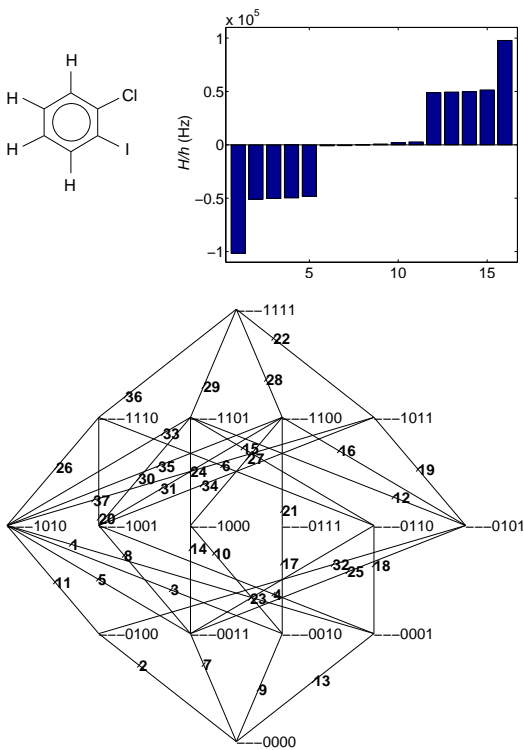}
\caption{\label{strhameltr} Molecular structure of 1-Chloro-2-iodobenzene
(top-left), the elements of diagonalized Hamiltonian (top-right),
and the energy level diagram with prominent transitions (bottom).
The eigenstates are labeled as shown.
}
\end{figure}
As the quantum register, 
we used the four $^1$H spins of 1-Chloro-2-iodobenzene (CIB; Figure \ref{strhameltr})
(purchased from Sigma Aldrich$^{\scriptsize\textregistered}$) oriented in 
liquid crystal ZLI-1132 (purchased from Merck$^{\scriptsize\textregistered}$)
forming a 10mM solution.  All experiments were carried out on a 500 MHz
Bruker Avance spectrometer at 300 K.  

\begin{figure}
\includegraphics[width=6cm,angle = -90]{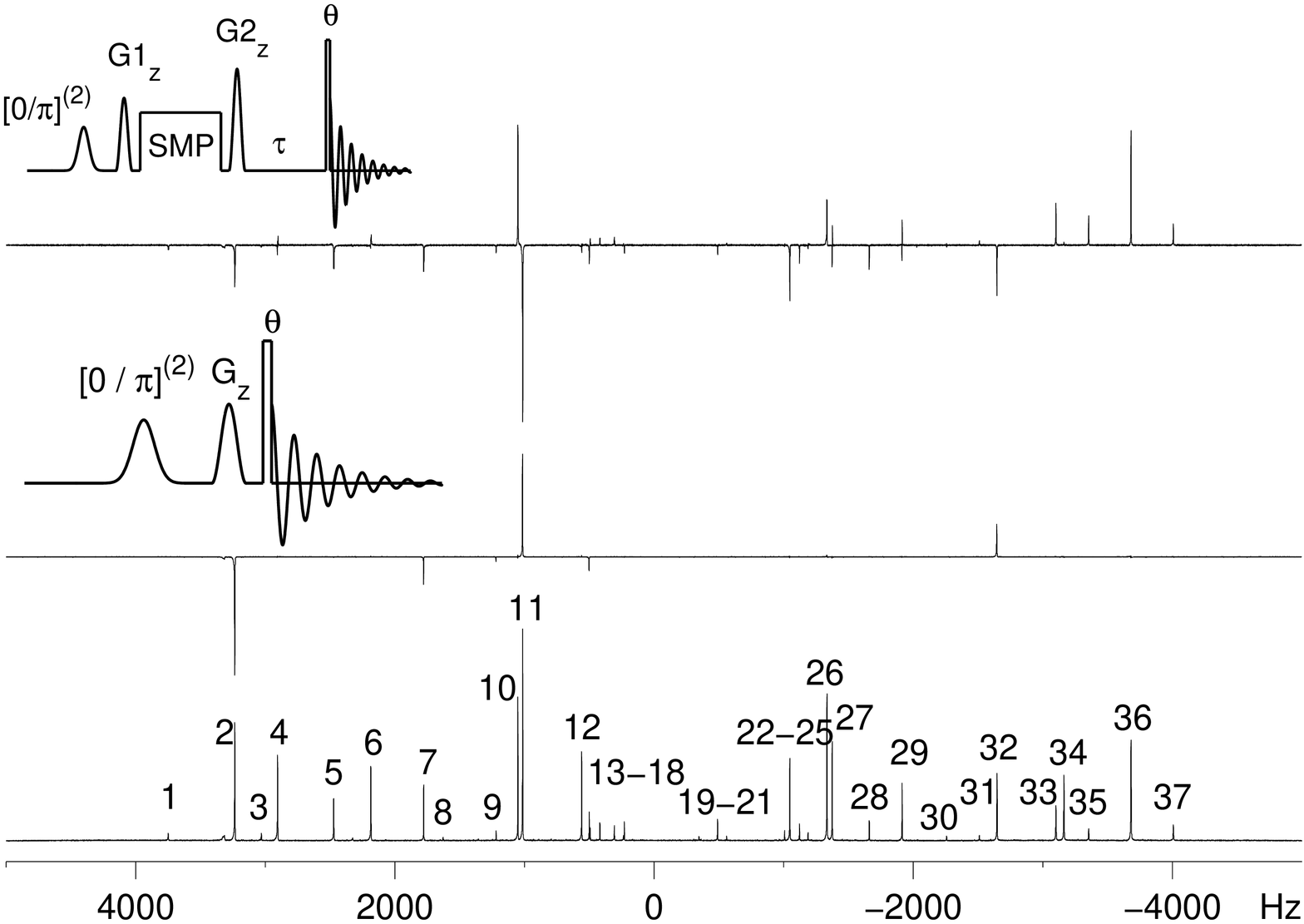}
\caption{\label{ngespec}
$^1$H spectra of CIB obtained by applying a small angle
pulse ($\theta = 3^\mathrm{o}$) to the equilibrium state (bottom),
$\vert 0100 \rangle \langle 0100 \vert -
 \vert 0000 \rangle \langle 0000 \vert$
POPS (middle), and
QNGE (top).  
The transitions numbers correspond to those in Figure \ref{strhameltr}.
The pulse sequences for POPS and QNGE
are shown in insets.  
(by applying a transition-selective Gaussian $\pi$ pulse 
of 10 ms duration)
Pulsed field gradients (G$_{\mathrm z}$) are used to
destroy coherence after POPS and after the SMP.
In the QNGE experiment, the
residual zero-quantum coherence
are de-phased by a 32-step average over a random delay $\tau$ (between 0 and 10 ms).
}
\end{figure}

Figure \ref{ngespec} shows the 
$^1$H NMR spectrum of the partially oriented CIB.  
The line widths of the various transitions
range from 1.7 Hz to 4.0 Hz indicating coherence times ($T_2^*$ relaxation times) 
between 1.8 s and 0.7 s.
The coherence times are sufficiently long to ignore
relaxation effects in the design of the SMP.

The procedure for analyzing the NMR spectra of partially oriented systems has been 
well studied \cite{diehlvol6,emsleybook}.  
We have developed a numerical 
procedure to iteratively determine the system Hamiltonian from 
its spectrum and a guess Hamiltonian \cite{maheshsamat}.  
The 37 strongest transitions of the CIB spectrum were used and a unique fit was obtained. 
The mean frequency and intensity errors 
between the experimental and the calculated spectra are less than 
0.1 Hz and 6\% respectively.
The elements of the diagonalized system Hamiltonian and 
the corresponding energy level diagram
are shown in Figure \ref{strhameltr}.

In NMR-QIP, the initial states are not pure states but are pseudopure states
that are isomorphic to pure states.  
The pseudopure states differ from the pure states by a uniform
background population on all states.  
It is easier, however, to prepare a pair of
pseudopure states (POPS) \cite{Fungpps}.  
We prepared the pair
$\vert 0100 \rangle \langle 0100 \vert - 
 \vert 0000 \rangle \langle 0000 \vert$.
The first term represents the desired initial state; the additional second part does not interfere with
the QNGE experiment,
because the operation IQFT, 
when acted on $\vert 00 \rangle$, creates a uniform superposition
of the ancilla qubits.  
Such a state is invariant under the $f(x) \oplus_{N_0}$ operation
and therefore the output state corresponding to $\vert 0000 \rangle \langle 0000 \vert$
is independent of the oracle $U_f$. 
The inset at the center of Figure \ref{ngespec} shows the pulse sequence
for preparing POPS and the corresponding spectrum.  The POPS spectrum
is obtained by linear detection pulse after inverting transition 2
and subtracting from the resulting spectrum the linearly detected
spectrum of the equilibrium state.

\begin{figure}
\includegraphics[height=7cm,angle = -90]{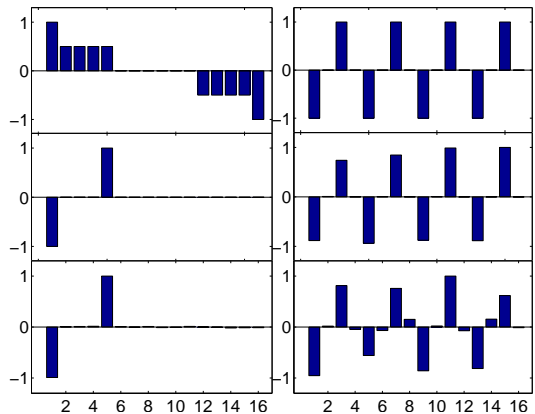}
\caption{\label{barplots}
Bar-plots showing the diagonal elements of the density matrix in the
eigenbasis: equilibrium (top left), simulated POPS
(middle left), experimental POPS (bottom left),
theoretical output of QNGE (top right),
simulated output (with the SMP shown in Figure \ref{smprob}; middle right),
and experimental output of the quantum algorithm (bottom right).}
\end{figure}

Since we use eigenbasis as the computational basis, the projective
measurement on to the computational basis is equivalent to 
measurement of populations after destroying the coherence.
First we dephase the coherence by using a pulsed field gradient (PFG).
A PFG does not efficiently dephase homonuclear
zero-quantum coherence.  Therefore we use a random delay 
(in between 0 and 10 ms) after the PFG and average over several (32)
transients.  The populations are then measured using a linear
detection (small flip-angle) pulse \cite{ernstbook}.
Using the normalization condition,
there are 15 unknowns for the 16 eigenstates.  The results of the diagonal-tomography
obtained by the mean of three sets each of 15 linearly independent transitions
are shown in Figure \ref{barplots}.

After the projective measurement in the eigenbasis at the end of the quantum algorithm,
the input qubits are in the state $\vert 10 \rangle$ encoding the gradient 
$\nabla f = 2$, 
while the ancilla qubits have equal probability in all possible states.  Therefore,
the theoretical output diagonal state of the combined system is 
$I_{an} \otimes (\vert 10 \rangle \langle 10 \vert -
\vert 00 \rangle \langle 00 \vert$), 
where $I_{an}$ is the identity operator for the ancilla qubits
and the two parts in the parenthesis correspond to the two parts 
of the POPS. 

The diagonal
correlation $C$ between the theoretical density matrix ($\rho_{\mathrm{T}}$) and the 
experimental density matrix ($\rho_{\mathrm{E}}$), is defined as
\begin{eqnarray}
C = \frac{ \mathrm{trace} [\vert \rho_{\mathrm{T}} \vert 
    \cdot \vert \rho_{\mathrm{E}}\vert]}
    {\sqrt{ \mathrm{trace} [\vert \rho_{\mathrm{T}} \vert ^2] 
    \cdot   \mathrm{trace} [\vert \rho_{\mathrm{E}} \vert ^2]}},
\end{eqnarray}
where $\vert \vert$ denotes extraction of the diagonal part.  The average
diagonal correlations were
0.999 and 0.979 for the POPS and the result of the QNGE, respectively.  
The lower value of the latter may be attributed to decoherence and
spectrometer nonlinearities.  

In conclusion, we have demonstrated quantum computation on the eigenbasis
of system Hamiltonian
using coherent control techniques.
As an example, we described the first implementation of
Jordan's algorithm for numerical gradient estimation.
Compared to the usual approach using weakly coupled systems,
the present method using strongly dipolar coupled systems
yields significantly faster execution times and is therefore less susceptible
to decoherence effects. 
From molecular and spectroscopic considerations a combination of homo
and heteronuclear spins in either liquid crystalline or molecular 
single crystalline environments is a natural way to build 
larger qubit-systems (with $>10$ qubits).
The coherent control of strongly dipolar coupled systems 
then becomes important and the present work is the
first step in this direction.
We believe that the coherent control techniques demonstrated here
for dipolar coupled nuclear spins will turn out to be essential also
for other solid-state implementations of quantum information processing.

\acknowledgments
This work was supported by the Alexander von Humboldt Foundation
and the DFG through grant numbers Su192/19-1 and Su192/11-1.

\end{document}